\documentclass[letterpaper,showpacs,aps,prd,10pt]{revtex4-2}

\usepackage[utf8]{inputenc}
\usepackage[english]{babel}

\usepackage{silence}
\usepackage{latexsym}
\usepackage{amssymb}
\usepackage{amsmath}
\usepackage{bm} % bold math
\usepackage{tabularx}
\usepackage{graphicx}
\usepackage{epsfig}
\usepackage{enumitem}
\usepackage{caption}
\usepackage{subcaption}

\usepackage[colorlinks=true, linkcolor=blue, citecolor=blue, urlcolor=blue]{hyperref}
\usepackage{comment}

\newcommand{\mathper}[1]{\text{\slshape #1}}

\usepackage{color}

\begin{document}

\captionsetup[figure]{labelfont=bf,singlelinecheck=off,justification=raggedright}

%%%%%%%%%%%%%%%%%
%%%   FRONT   %%%
%%%%%%%%%%%%%%%%%

\title{Dynamical Evolutions of Electrically Charged Proca Stars}

\author{Yahir Mio}
\email{jorge.mio@correo.nucleares.unam.mx}

\author{Miguel Alcubierre}
\email{malcubi@nucleares.unam.mx}

\affiliation{Instituto de Ciencias Nucleares, Universidad Nacional
Aut\'onoma de M\'exico, Circuito exterior C.U., A.P. 70-543, Ciudad de M\'exico 04510, M\'exico.}

%%%%%%%%%%%%%%%%
%%%   DATE   %%%
%%%%%%%%%%%%%%%%

\date{\today}

%%%%%%%%%%%%%%%%
%%%   PACS   %%%
%%%%%%%%%%%%%%%%


%%%%%%%%%%%%%%%%%%%%
%%%   ABSTRACT   %%%
%%%%%%%%%%%%%%%%%%%%

\begin{abstract}
In a previous work we constructed different families of stationary electrically charged Proca stars characterized by a charge parameter $q$, by solving the Einstein--Maxwell--Proca system in spherical symmetry, and imposing a harmonic time dependence ansatz for the Proca field \cite{Mio:2025}. We showed that there is a critical value for the charge $q_c$ that corresponds to the value for which the Coulomb repulsion of the charged Proca field exactly cancels the Newtonian gravitational attraction, and we found that supercritical solutions can only exist for a limited range of charges above this critical value $q>q_c$. Here we study the dynamical stability properties of these charged Proca stars by adding a small but finite perturbation to the original stationary configurations, and then performing numerical evolutions while keeping the spherical symmetry. We show that, for any given family, the parameter space can be separated into three regions corresponding to gravitationally bound stable configurations, gravitationally bound unstable configurations, and gravitationally unbound unstable configurations.  For the unstable configurations we follow the evolution in time in order to determine their final state, and find that this final state can be collapse to a charged Reissner--Nordstr\"om black hole, migration to a new state in the stable branch, or dispersion to infinity, depending on the value of the binding energy and the specific form of the perturbation.
\end{abstract}

%%%%%%%%%%%%%%%%%%%%%%
%%%   MAKE TITLE   %%%
%%%%%%%%%%%%%%%%%%%%%%

\maketitle

%%%%%%%%%%%%%%%%%%%%%%%%
%%%   INTRODUCTION   %%%
%%%%%%%%%%%%%%%%%%%%%%%%

\section{Introduction}

First introduced by Brito in \cite{Brito:2015pxa}, Proca stars have become one of the most widely studied exotic compact objects (ECO's) in recent years~\cite{SanchisGual:2017, Aoki:2022mdn, Aoki:2022woy, Hernandez:2023tig, Zhang:2025a,Lazarte:2024jyr,Lazarte:2025wlw}. Such objects consist of self-gravitating solutions for a massive complex vector boson field (the Proca field) minimally coupled to gravity, for which the vector field oscillates harmonically in time while the spacetime geometry remains static. Several of its most remarkable properties of these objects have been explored in the literature (see~\cite{Herdeiro:2020jzx} for a review), and studies of the dynamical evolution have established that this family of solutions admits an important stable branch, a crucial requirement for the physical viability of such objects. The scientific literature also contains several studies of compact objects formed by Proca fields coupled to additional matter fields~\cite{Ma:2023bhb, Ma:2023vfa, Herdeiro:2023a, Herdeiro:2024a, Herdeiro:2024b, Jockel:2023rrm, Pombo:2023sih, Herdeiro:2023a, Herdeiro:2024b}. Furthermore, numerical simulations of Proca star collisions have indicated that this phenomenon is compatible with the GW190521 gravitational wave event~\cite{CalderonBustillo:2020fyi, Luna:2024kof}, initially classified as a collision of black holes~\cite{Abbott:2020}. For these reasons, Proca stars are regarded as possible black hole mimickers~\cite{Rosa_2022,Sengo_2024}, and have attracted the interest of astrophysicists (detailed studies of Proca star collisions and the subsequent emission of gravitational waves can be found in~\cite{Luna:2024kof, SanchisGual:2019,Sanchis-Gual:2022mkk}).

In~\cite{SalazarLandea:2016bys}, electrically charged Proca stars (CPS) were first introduced. Such configurations can be defined as exotic compact objects formed by massive complex charged vector boson fields (i.e. Proca fields). In general, CPSs are characterized by two parameters: the field charge $q$ and the central scalar potential $\varphi_0$. If we consider a family of solutions sharing a fixed value of $q$, one finds that all such families exhibit a behavior qualitatively similar to that of uncharged $q=0$ Proca stars. In addition, a critical value of the charge $q_c$ can be identified that corresponds to the value for which the Coulomb repulsion of the charged Proca field exactly cancels the Newtonian gravitational attraction, beyond which no solutions should be expected. However, when we extended the study of CPS in~\cite{Mio:2025}, we were able to find solutions for a limited range of values such that $q>q_c$. It is worth mentioning that such supercritical solutions were also found in the case of charged boson stars in~\cite{Pugliese_2013,Lopez:2023phk}. In our work, we also made evident the repulsive role of the electric charge in the star. To date, studies of CPSs have been restricted to the static, spherically symmetric case. This is worth emphasizing, as such spherical configurations may not correspond to the true ground state of the system. Indeed, for the uncharged case, it was shown in~\cite{Herdeiro:2024a} that spherically symmetric Proca stars are in fact unstable and dynamically evolve toward non-spherical prolate configurations.

Since the stability of any compact object is of fundamental importance, here we perform a numerical study of the dynamical evolution of CPSs while maintaining the spherical symmetry, using the solutions obtained in~\cite{Mio:2025} as initial data. We have evolved both perturbed and unperturbed initial data, where the perturbations are achieved by adding or subtracting energy at the star's center in a self-consistent way. As expected, we found that the different families of charged configurations exhibit a behavior similar to the uncharged case. By exploring the parameter, we identify a region consisting of stable configurations. However, we also identify a region of unstable solutions that exhibit different outcomes under dynamical evolution: gravitationally bound but unstable stars can migrate to a stable star or collapse to a Reisner-Nordstr\"om black hole, while unstable unbound stars can either collapse or disperse. The final outcome depends not only on the type of star, but also on the nature of the applied perturbation.

This article is organized as follows: In Section II we describe the Einstein-Maxwell-Proca system of equations and focus on the case of spherical symmetry in the 3+1 formalism. In Section III we describe the construction of the initial data and the perturbation scheme adopted. Next, in Section IV, we explain how we find the total charge, total mass, and binding energy of our configurations. In Section V we describe how we find the apparent horizon and its associated mass. The numerical results are presented in Section VI. Finally, in Section VII we present our conclusions.

%%%%%%%%%%%%%%%%%%%%%%%%%%%%%%%%%%%%%%%%%
%%%   EINSTEIN-MAXWELL-PROCA SYSTEM   %%%
%%%%%%%%%%%%%%%%%%%%%%%%%%%%%%%%%%%%%%%%%

\section{The Einstein-Maxwell-Proca system}

As described in detail in our previous paper~\cite{Mio:2025}, we consider an electrically charged complex Proca field $\mathcal{W}_{\mu\nu}$ with mass parameter $m$, coupled to an electromagnetic field $F_{\mu\nu}$ through a charge parameter $q$ which plays the role of the coupling constant. The Proca and Maxwell field tensors $\mathcal{W}_{\mu\nu}$ and $F_{\mu\nu}$ are given in terms of their respective  potential 1-forms $X_\mu$ and $A_\mu$  as:
\begin{align}
\mathcal{W}_{\mu\nu} &= \mathcal{D}_\mu X_\nu - \mathcal{D}_\nu X_\mu  \; ,\\
F_{\mu\nu} &= \nabla_\mu A_\nu - \nabla_\nu A_\mu \; ,
\end{align}
where we have introduced the gauge-invariant derivative $\mathcal{D}_\mu:=\nabla_\mu-iqA_\mu$. We consider both fields minimally coupled to the gravitational field, leading to the Einstein-Maxwell-Proca (EMP) system of equations.

We will use the 3+1 formalism of General Relativity, where spacetime is foliated into spacelike hypersurfaces with a timelike future-pointing normal vector $n^\mu$ (for a detailed introduction into the 3+1 formalism see e.g.~\cite{Alcubierre08a}).  The projector operator onto the spatial hypersurfaces is then defined as $\gamma^\mu_\nu := \delta^\mu_\nu + n^\mu n_\nu$; its spacelike covariant components $\gamma_{ij}$ (with $i,j=1,2,3$) play the role of the three-dimensional induced metric tensor on the spatial hypersurfaces. The normal vector $n^\mu$ and the projector operator $\gamma^\mu_\nu$ can be used to decompose 4-vectors and tensors into their spatial and time components. Thus, we can decompose the potential 1-forms as:
\begin{align}
\varphi &:= - n^\nu X_\nu \; , &
x_\mu &:= \gamma_\mu^\nu X_\nu \; , \label{3} \\
\phi &:= - n^\nu A_\nu \; , &
a_\mu &:= \gamma_\mu^\nu A_\nu \; .
\end{align}
and the fields as:
\begin{align}
\mathcal{E}_\mu &:= n^\nu \mathcal{W}_{\mu\nu} \; , &
\mathcal{B}_\mu &:= n^\nu \mathcal{W}^*_{\mu\nu} \; , \\
E_\mu &:= n^\nu F_{\mu\nu} \; , &
B_\mu &:= n^\nu F^*_{\mu\nu} \; , \label{6}
\end{align}
where $\mathcal{W}^*_{\mu\nu} := - \frac{1}{2} \: \varepsilon_{\alpha\beta \mu\nu} \mathcal{W}^{\mu\nu}$ and $F^*_{\mu\nu}=- \frac{1}{2} \: \varepsilon_{\alpha\beta \mu\nu} F^{\mu\nu}$ are the dual field tensors, with $\varepsilon_{\alpha\beta\mu\nu}$ the Levi--Civita tensor which is in turn defined in terms of the Levi--Civita symbol $\epsilon_{\alpha\beta\mu\nu}$ and the determinant of the spacetime metric tensor $g$ as  $\varepsilon_{\alpha\beta\mu\nu} := \sqrt{-g} \: \epsilon_{\alpha\beta\mu\nu}$ (we use a convention such that $\epsilon_{0123} = 1 = - \epsilon^{0123}$). 

We can clearly see that $(\varphi,\phi)$ are scalar potentials, while $(x_\mu,a_\mu)$ are vector potentials. In addition, the latter together with $(\mathcal{E}_\mu, E_\mu, \mathcal{B}_\mu, B_\mu)$ are purely spatial vectors in the sense that $n^\mu x_\mu = n^\mu a_\mu = 0$, $n^\mu E_\mu = n^\mu \mathcal{E}_\mu = 0$, and $n^\mu B_\mu = n^\mu \mathcal{B}_\mu = 0$.  The vectors $(\mathcal{E}_\mu, E_\mu)$ and $(\mathcal{B}_\mu, B_\mu)$ play the role of electric and magnetic fields, respectively, for both the Proca and Maxwell cases.

We will be interested in spherically symmetric Proca stars. Therefore, we consider a metric whose line element is given in spherical coordinates $(r,\theta,\Phi)$ by (do not confuse the azimuthal angle $\Phi$ with the Maxwell scalar potential $\phi$):
\begin{equation}
\label{eq:4metric}
ds^2 = - \alpha^2 dt^2 + \Psi^4 \left( \mathper{a} \: dr^2 + r^2 \mathper{b} \: d\Omega^2 \right) \; ,
\end{equation}
where ($\alpha, \Psi, \mathper{a}, \mathper{b}$) are functions of $(r,t)$ only, and $d\Omega^2 = d \theta^2 + \sin^2 \theta \: d \Phi^2$ is the usual solid angle element.  Notice that for simplicity we are assuming that the shift vector vanishes. The metric above might seem too general because of the inclusion of a conformal factor $\Psi$ which is not really needed for a generic spherically symmetric spacetime. Indeed, for our initial data we always take take $\Psi=\mathper{b}=1$ (see below). However, this form of the metric is required for our dynamical simulations since we will be using the Baungarte-Shapiro--Shibata-Nakamura (BSSN) formulation for the evolution equations~(see e.g. \cite{Baumgarte:1998te,Shibata95}).

Using the definitions given in (\ref{3}-\ref{6}) and the metric given in~(\ref{eq:4metric}), the evolution equations for the Proca and Maxwell potentials and fields take the form (see~\cite{Mio:2025} for a detailed derivation):
\begin{align}
\partial_t \varphi &= - \frac{x_r}{\mathper{a} \Psi^4} \: \partial_r\alpha  - \frac{\alpha}{\mathper{a} \Psi^4} \left[\partial_rx_r +x_r \left(\frac{2}{r} - \frac{\partial_r \mathper{a}}{2\mathper{a}} + \frac{\partial_r \mathper{b}}{\mathper{b}} + 2\frac{\partial_r \Psi}{\Psi} \right)\right] + \alpha \varphi K \notag \\
&+ iq\alpha \left(-\phi\varphi + \frac{1}{\mathper{a} \Psi^4} a_rx_r -\frac{\mathper{a} \Psi^4}{m^2} E^r \mathcal{E}^r\right)
\; , \label{ev-varphi-3+1} \\
\partial_t \phi &=  - \frac{a_r}{\mathper{a} \Psi^4} \: \partial_r\alpha - \frac{\alpha}{\mathper{a} \Psi^4} \left[\partial_ra_r +a_r \left(\frac{2}{r} - \frac{\partial_r \mathper{a}}{2\mathper{a}} + \frac{\partial_r \mathper{b}}{\mathper{b}} + 2 \frac{\partial_r \Psi}{\Psi^4} \right)\right] + \alpha \phi K \; ,
\label{ev-phi-3+1} \\
\partial_t x_r &= - \alpha \left( \mathper{a} \Psi^4 \mathcal{E}^r
+ \partial_r \varphi \right) - \varphi \partial_r\alpha - iq \alpha \left( x_r \phi - a_r \varphi \right) \; ,
\label{ev-x-3+1} \\
\partial_t a_r &= - \alpha \left( \mathper{a} \Psi^4 E^r + \partial_r\phi \right)
- \phi \partial_r \alpha \; ,
\label{ev-a-3+1} \\
\partial_t \mathcal{E}^r &= + \alpha \left( K \mathcal{E}^r
+ \frac{m^2}{\mathper{a} \Psi^4} \: x_r \right) - iq \alpha \phi \mathcal{E}^r \; ,
\label{ev-Ep-3+1} \\
\partial_t E^r &= + \alpha K E^r - 4 \pi \alpha j_Q^r \; , \label{ev-Em-3+1}
\end{align}
with $K=K^m_m$ the trace of the extrinsic curvature of the spatial hypersurfaces, and where $(x_r,a_r,\mathcal{E}^r, E^r)$ are the radial components of the spatial potential 1-forms and electric field vectors respectively. Notice in particular that in spherical symmetry the magnetic fields vanish, $\mathcal{B}_\mu = B_\mu = 0$.

In the same way, the electric (Gauss) constraints take the form:
\begin{align}
\partial_r \mathcal{E}^r + \mathcal{E}^r \left(\frac{2}{r} + \frac{\partial_r \mathper{a}}{2\mathper{a}} + \frac{\partial_r \mathper{b}}{\mathper{b}} + 6 \frac{\partial_r \Psi}{\Psi} \right) &= - m^2\varphi + iqa_r\mathcal{E}^r \; ,
\label{cons-Ep-3+1} \\
\partial_r E^r + E^r \left(\frac{2}{r} + \frac{\partial_r \mathper{a}}{2\mathper{a}}
+ \frac{\partial_r \mathper{b}}{\mathper{b}} + 6 \frac{\partial_r \Psi}{\Psi} s\right)
&= 4 \pi \rho_Q \; ,
\label{cons-Em-3+1}
\end{align}
with $\rho_Q$ the charge density associated with the Proca field given by:
\begin{equation}
\rho_Q = \frac{iq}{8\pi} \left( x_r \bar{\mathcal{E}}^r - \bar{x}_r \mathcal{E}^r \right) \; , \label{charge-density-EMP}
\end{equation}
where the overbar denotes the complex conjugate of the corresponding quantity. Additionally, we can also define the radial component of the charge current density $j^r_Q$, which takes the form:
\begin{equation}
j^r_Q = \frac{iq}{8\pi} \left( \varphi \bar{\mathcal{E}}^r - \bar{\varphi} \mathcal{E}^r \right) \; .
\end{equation}

\vspace{5mm}

To complete the system we must consider the Einstein field equations $G_{\mu\nu}=8\pi T_{\mu\nu}$, with $T_{\mu\nu} $ the stress-energy tensor for the Proca and Maxwell fields (we use units such that $G=c=\hbar=1)$.  From the stress-energy tensor we can define the energy density $\rho := n^\mu n^\nu T_{\mu\nu}$, the momentum density $J^i := -n^\mu \gamma^{r\nu} T_{\mu\nu}$, and the spatial stress tensor $S_{ij} := \gamma_i^\mu \gamma_j^\nu T_{\mu\nu}$.  In our particular case these quantities reduce to:
\begin{align}
\rho:=  &= \frac{1}{8\pi} \left[\mathcal{E}_r \bar{\mathcal{E}}^r + m^2(\varphi\bar{\varphi} + x_r\bar{x}^r) + E_rE^r\right] \; ,
\label{eq:energy-density} \\
J^r := &= \frac{1}{8\pi} m^2\left(x^r\bar{\varphi} + \bar{x}^r \varphi\right) \; ,
\label{eq:momentum-density} \\
S_{ij} :=  &= \frac{1}{8\pi} \left\{\gamma_{ij} \left(\mathcal{E}_r \bar{\mathcal{E}}^r + E_r E^r\right)\right. - \left(\mathcal{E}_i \bar{\mathcal{E}}_j + \bar{\mathcal{E}}_i \mathcal{E}_j  + 2 E_i E_j\right) \notag \\
& \left. + m^2 \left[ \left(x_i \bar{x}_j + \bar{x}_i x_j \right) - \gamma_{ij} \left(x_r \bar{x}^r - \varphi \bar{\varphi} \right)\right] \rule{0mm}{4mm} \right\} \; .
\end{align}
Notice that the momentum density only has a non-zero radial component $J^r$, while the stress tensor $S_{ij}$ has non-zero diagonal components given by $S_{rr}$, $S_{\theta \theta}$ and $S_{\Phi \Phi} = \sin^2 \theta \: S_{\theta \theta}$.

As mentioned above, the spacetime variables are evolved using the BSSN formulation of general relativity adapted to spherical symmetry~\cite{Brown:2009dd,Alcubierre:2010is}.  We will not write explicitly those evolution equations here since they can be found in detail in reference~\cite{Alcubierre:2010is}.

%%%%%%%%%%%%%%%%%%%%%%%%%%%%%%%%%%%%
%%%   STATIONARY INITIAL DATA   %%%%
%%%%%%%%%%%%%%%%%%%%%%%%%%%%%%%%%%%%

\section{Initial data}

\subsection{Stationary initial data} 

For constructing stationary charged Proca stars we consider the following ansatz for the Proca potentials:
\begin{equation}
\varphi(t,r) = \varphi(r)e^{-i\omega t} \; , \qquad
x_r(t,r) = ix(r)e^{-i\omega t} \; . \label{ansatz-pot}
\end{equation}
with $\varphi(r)$ and $x(r)$ real functions, and $\omega$ a real frequency. This form of the Proca potentials guarantees that the quantities associated with the stress-energy tensor and the charge density $\rho_Q$ are time independent, and also that the current density and momentum density vanish, $J^r = j^r_Q=0$.  We then have an electrostatics problem for which the Maxwell potentials take the simple form:
\begin{equation} \label{pot_max}
\phi(t,r) = \phi(r) \; , \qquad
a_r(t,r) = 0 \; .
\end{equation}
In turn, our previous assumptions also imply that the Maxwell and Proca electric fields
take the form:
\begin{equation}
E^r(t,r) = E(r) \; , \qquad
\mathcal{E}^r(t,r) = \mathcal{E}(r)e^{-i\omega t} \; .
\label{fields-E}
\end{equation}
Finally, since our spacetime is static the extrinsic curvature will also vanish, that is $K_{ij}=0$ and hence $K=0$.

For stationary solutions we take the conformal factor to be $\Psi=1$. Additionally, we choose to work with the areal radius so that we also have $\mathper{b}=1$.  We then find the following system of radial ordinary differential equations:
\begin{align}
\psi' &= - \mathper{a} \mathcal{E} - \frac{(\omega - q\vartheta)x} {\alpha} \; ,
\label{eq-rad-psi} \\
x' &= \frac{(\omega - q\vartheta) \psi \mathper{a}}{\alpha^2} - x \left[ \frac{1}{r} \left( \mathper{a} + 1 \right) + 4 \pi r\mathper{a} \left( S^r_r - \rho \right) \right] - q \: \frac{\mathper{a}^2 E \mathcal{E}}{m^2} \; ,
\label{eq-rad-x} \\
\vartheta' &= -\alpha \mathper{a}E \; , \label{eq-rad-vartheta} \\
E' &= - q x \mathcal{E} - E \left[ \frac{1}{2r}(5-\mathper{a}) + 4 \pi r \mathper{a} \rho \right] \; ,
\label{eq-rad-E} \\
\mathcal{E}' &= - \dfrac{m^2 \psi}{\alpha} - \mathcal{E} \left[ \frac{1}{2r}(5-\mathper{a}) + 4 \pi r \mathper{a} \rho \right] \; ,
\label{eq-rad-Ep}
\end{align}
where the prime indicates derivatives with respect to $r$, and where we have introduced the change of variables $\psi := \alpha \varphi$ and $\vartheta := \alpha \phi$. The first three equations above come directly from the evolution equations for $x$, $\varphi$ and $a$; equations~\eqref{ev-x-3+1}, \eqref{ev-varphi-3+1} and \eqref{ev-a-3+1}, respectively.  The evolution equation for $\phi$, eq.~\eqref{ev-phi-3+1}, is now trivially satisfied, as is the evolution equation for $E^r$, eq.~\eqref{ev-Em-3+1}.  On the other hand, the last two equations above for $E'$ and $\mathcal{E}'$ come from the Gauss constraints~\eqref{cons-Ep-3+1} and~\eqref{cons-Em-3+1}.  Here it is important to mention the fact that, from the evolution equation for $\mathcal{E}^r$, eq.~\eqref{ev-Ep-3+1}, one can in fact solve explicitly for $\mathcal{E}$ to obtain:
\begin{equation} \label{eq-Ep-analitic}
\mathcal{E} = - \frac{\alpha m^2 x}{\mathper{a}(\omega - q\alpha\phi)} \; .
\end{equation}
This expression is fully consistent with the system of equations above, as can be easily seen by direct substitution, so it can be used instead of equation~\eqref{eq-rad-Ep}.  However, this is only true for stationary solutions, and not for the perturbed initial data that we will consider in the next section, where we do need to use equation~\eqref{eq-rad-Ep}.

In addition, we have the following equations for the radial metric component $\mathper{a}(r)$ and the lapse function $\alpha(r)$:
\begin{align}
\mathper{a}' &= \mathper{a} \left[\frac{1}{r}(1-\mathper{a}) + 8\pi r \mathper{a} \rho\right] \; , \label{eq-rad-a} \\
\alpha' &= \alpha \left[\frac{1}{2r}(\mathper{a}-1) + 4 \pi r \mathper{a} S^r_{\phantom{r}r}\right] \; . \label{eq-rad-alpha}
\end{align}
The equation for $\mathper{a}'$ arises directly from the Hamiltonian constraint, while the equation for $\alpha'$ comes from the fact that we are using the areal radial coordinate, so that $b=1$ and $\partial_t b=0$.

Finally, for the energy density and radial component of the stress tensor $S^r_r$ we have:
\begin{align}
\rho &= + \frac{1}{8\pi} \left[ \mathper{a} \mathcal{E}^2 + m^2 \left(\varphi^2 + \frac{x^2}{\mathper{a}}\right) + \mathper{a} E^2\right] \; ,
\label{rho-epc} \\
S^r_r &= -\frac{1}{8\pi} \left[ \mathper{a} \mathcal{E}^2 - m^2 \left(\varphi^2 + \frac{x^2}{\mathper{a}}\right) + \mathper{a} E^2 \right] \; . \label{s-epc}
\end{align}
Notice that these two expressions are quite similar, with only some sign differences.

Since charged Proca stars are compact and regular at the origin, the radial equations must satisfy a series of boundary conditions. In particular, at infinity the following must be true:
\begin{equation} \label{cond-inf}
\alpha, \mathper{a} \rightarrow 1 \; , \qquad
\varphi, x, \phi, E \rightarrow 0 \; ,
\end{equation}
while at the origin we must have:
\begin{align}
\mathper{a}(0) &= 1 \; , &
\alpha(0) &= \alpha_0 \; , &
\varphi(0) &= \varphi_0 \; , &
\phi(0) &=\phi_0 \; , &
x(0) &= 0 \; , &
E(0) &= 0 \; , \\
\mathper{a}'(0) &= 0 \; , &
\alpha'(0) &= 0 \; , &
\varphi'(0) &= 0 \; , &
\phi'(0) &= 0 \; , &
x'(0) &= \frac{\omega \varphi_0}{3\alpha_0} \; , &
E'(0) &= 0 \; ,
\end{align}
with $\alpha_0$, $\varphi_0$ and $\phi_0$ positive real constants. Our system of equations has in fact only two degrees of freedom, namely $q$ and $\varphi_0$ (and in principle $m$, but we always take $m=1$ since solutions for any other value of $m$ can be found by rescaling).  The constants $\alpha_0$ and $\phi_0$, while not given a priori, are fixed by the boundary conditions at infinity.  In practice, we initially take $\alpha_0=1$ and $\phi_0=0$, and later determine their correct values using a rescaling and a gauge transformation in order to obtain $\alpha \rightarrow 1$ and $\phi \rightarrow 0$ at infinity.  The frequency $\omega$, on the other hand, is found by choosing values for $q$ and $\varphi_0$ and then solving an eigenvalue problem in order to guarantee exponential decay of the Proca field at infinity. For more details on the numerical methods and algorithms used to solve the system see~\cite{Mio:2025}.

Families of solutions corresponding to stationary charged Proca stars for different values of the charge $q$ have already been presented in our previous work~\cite{Mio:2025}, where it was shown that solutions exist for arbitrary values of $\varphi_0>0$ as long as the charge is smaller than a critical value $q_c=m$.  We also found solutions for slightly supercritical values of  the charge such that $1 < q/m \lesssim 1.04$, but only for a narrow range of values of $\varphi_0$.

%%%%%%%%%%%%%%%%%%%%%%%%%%%%%%%%%%%
%%%   PERTURBED INITIAL DATA   %%%%
%%%%%%%%%%%%%%%%%%%%%%%%%%%%%%%%%%%

\subsection{Perturbed initial data}

In the previous section we considered the case of stationary Proca star configurations. In all such configurations the complex Proca field oscillates harmonically, while the Maxwell field and spacetime variables remain static. These solutions can be evolved in time using our system of evolution equations~\eqref{ev-varphi-3+1}-\eqref{ev-Em-3+1}, and ideally should remain in a stationary state forever. In practice, however, when they are evolved numerically this is no longer true since the numerical truncation error introduces a perturbation in itself. In such a case, stable solutions should show small oscillations around the stationary solution that do not grow with time, while unstable solutions will rapidly move away from the stationary solution and either collapse to a black hole, disperse to infinity, or migrate to a stable solution. 

However, perturbing solutions only with numerical truncation error has two serious disadvantages: first, the specific form of the truncation error can not be controlled in any detail; and second, as the resolution is increased this truncation error becomes smaller (as should be expected), which makes it difficult to study the numerical convergence of our solutions. Because of this we prefer to consider initial data with a finite self-consistent perturbation. 

In order to construct our perturbed initial data we then first consider a specific stationary solution for given values of $q$ and $\varphi_0$. Notice that this involves solving for the seven functions $(\varphi,x,\phi,E,\mathcal{E},\mathper{a},\alpha)$, and finding a specific value for the frequency $\omega$. We then add a perturbation to the Proca scalar potential $\varphi$ at $t=0$ of the form $\varphi_{\text{pert}}(r) = \varphi(r) + \delta \varphi(r)$, and leave the vector potential $x(r)$ unchanged. Notice that, since we only perturb the real part of the Proca scalar potential while leaving the vector potential unchanged and purely imaginary at $t=0$, the momentum density $J^r$ given by equation~\eqref{eq:momentum-density} remains equal to zero initially, so that the momentum constraint is still trivially satisfied.  Also, the Proca electric field $\mathcal{E}^r$ remains purely real at $t=0$, guaranteeing that the current density $j^r_Q$ also vanishes initially.

Specifically, we add a Gaussian perturbation to $\varphi(r)$ centered around the origin of the form: 
\begin{equation}
\delta \varphi (r) = \epsilon \varphi_0 \exp \left( - r^2 / \sigma^2 \right) \; ,
\end{equation}
where $\epsilon$ is a small dimensionless parameter that controls the amplitude of the perturbation, while $\sigma$ determines its width. Notice that since the amplitude of the Gaussian above is given by $\epsilon \varphi_0$, the parameter $\epsilon$ represents a fraction of the initial unperturbed amplitude $\varphi_0$.
Once we have added a perturbation to the Proca scalar potential $\varphi(r)$, while leaving the vector potential $x(r)$ unchanged, we now need to solve again for the five quantities $(\vartheta,E,\mathcal{E},\mathper{a},\alpha)$ using equations~\eqref{eq-rad-vartheta}-\eqref{eq-rad-Ep} and~\eqref{eq-rad-a}-\eqref{eq-rad-alpha}. This procedure then provides us with fully consistent perturbed initial data that satisfies the Gauss and Einstein constraints.

%%%%%%%%%%%%%%%%%%%%%%%%%%%%%%%%%%%%%%%%%%%
%%%   MASS, CHARGE AND BINDING ENERGY   %%%
%%%%%%%%%%%%%%%%%%%%%%%%%%%%%%%%%%%%%%%%%%%

\section{Charge, Total mass, and binding energy}

There are several quantities that one can calculate in order to better characterize the  properties of our charged Proca star configurations.  Consider first the total electric charge $Q$, which is given by the volume integral over the physical three-dimensional space of the charge density $\rho_Q$ as:
\begin{equation}
Q := 4 \pi \int_0^\infty \rho_Q \left( \mathper{a}^{1/2} \mathper{b}
\Psi^6 r^2 \right) dr \; ,
\label{eq:total-charge}
\end{equation}
where the term in parentheses comes from the square root of the determinant of the spatial metric, $\gamma = \mathper{a} \mathper{b}^2 \psi^{12} r^4$. The charge density $\rho_Q$ is given in general by equation~(\ref{charge-density-EMP}). Since our configurations are such that at $t=0$ the Proca electric field $\mathcal{E}$ is purely real while the Proca vector potential $x$ is purely imaginary, this reduces initially to:
\begin{equation}
\rho_Q = -\frac{q}{4\pi} \: x \mathcal{E} \; .
\end{equation}
However, this will no longer be true during a time evolution, so that for $t>0$ we need to use the general expression~(\ref{charge-density-EMP}). In addition, if we consider that the Proca star consists of $N$ particles, each with charge $q$, then we can define the total number of particles as $N:=Q/q$.

\vspace{5mm}

%%%%%%%%%%%%%%%%%%%%%%%%%%%%%%%%%%%%%%%%%%

The total mass $M$ of our Proca stars can be determined in two different ways. One can start from the Hamiltonian constraint:
\begin{equation} \label{hamiltonian-constrain}
R^{(3)} = 16 \pi \rho - K^2 + K_{ij} K^{ij} \; ,
\end{equation}
where $R^{(3)}$ is the three-dimensional Ricci scalar (related only to the spatial metric). 
Now, for a spherically symmetric spacetime one can show that the spatial metric can always be written in terms of the areal radius $r_a$ as:
\begin{equation}
dl^2 = \frac{dr_a^2}{1 - 2 m(r_a)/r_a} + r_a^2 d \Omega^2 \; ,
\end{equation}
with $m(r_a)$ the so-called Misner--Sharp mass function.  Notice that for compact objects the Misner--Sharp mass coincides with the total mass ADM of the spacetime asymptotically, as can be seen by comparing the expression above with the Schwarzschild metric.

When the metric is written as above, the Ricci scalar $R^{(3)}$ turns out to be given simply by:
\begin{equation} \label{Ricci-ms}
R^{(3)} = \left( \frac{4}{r_a^2} \right) \dfrac{dm(r_a)}{dr_a} \; .
\end{equation}
By equating (\ref{hamiltonian-constrain}) and (\ref{Ricci-ms}) we obtain an expression for $dm(r_a)/dr_a$, which we can then integrate to find that the total mass as:
\begin{equation}
M_{int} = \int_0^\infty \left[ 4\pi\rho + \frac{1}{4} (K_{ij} K^{ij} - K^2)\right] r_a^2 dr_a \: .
\end{equation}
The integral above is to be calculated in terms of the areal radius $r_a$.  However, the form of the spatial metric used in our numerical code is given by:
\begin{equation} \label{eq:3metric}
dl^2 = \Psi^4 \left( \mathper{a} \: dr^2 + r^2 \mathper{b} \: d\Omega^2 \right) \; ,
\end{equation}
so that in general the radial coordinate $r$ does not coincide with the areal radius $r_a$, except at $t=0$. The relation between the coordinate radius and the area radius is given in general by $r_a = \Psi^2 \mathper{b}^{1/2} r$, so that:
\begin{equation}
r_a^2 dr_a =r^2 \Psi^6 \mathper{b}^{3/2} \left[ 1 + r \left( \dfrac{\partial_r \mathper{b}}{2\mathper{b}} + 2 \: \dfrac{\partial_r \Psi}{\Psi} \right) \right] dr \; .
\end{equation}
The mass integral then takes the following general form in terms of the coordinate radius $r$:
\begin{equation} \label{masa-integr}
M_{int} = \int_0^\infty \left[ 4\pi\rho + \frac{1}{4} (K_{ij} K^{ij} - K^2)\right] \left[1+ r \left( \dfrac{\partial_r \mathper{b}}{2\mathper{b}} + 2 \dfrac{\partial_r \Psi}{\Psi} \right) \right] r^2 \Psi^6 \mathper{b}^{3/2} dr \; .
\end{equation}
For momentarily static initial data such that $K_{ij}=0$, and taking $\Psi=\mathper{b}=1$ (which our configurations satisfy at $t=0$, but not later), this reduces simply to:
\begin{equation}
M_{int} =  4 \pi \int_0^\infty \rho r^2 dr \; .
\end{equation}
It is interesting to compare this last expression with that for the total charge $Q$, equation~\eqref{eq:total-charge} above.  Even taking $\Psi=\mathper{b}=1$ in the expression for the total charge, we notice that the charge integral still contains a factor of $\mathper{a}^{1/2}$ coming from the volume element that is not present in the mass integral.  This implies that, while the charge integral is taken over the physical curved volume element, the mass integral is taken over the flat volume element. The reason for this is that the mass integral is not just the integral of the energy density over physical space, but must also take into account the effect of the (negative) potential energy.

Another way to calculate the total mass of the charged Proca stars is from the fact that we know that the Reissner--Nordstr\"om solution is the only static electro-vacuum solution in spherical symmetry. Then we can assume that, since the Proca field of our configurations decays exponentially, far from the center of the stars the metric should rapidly converge to the Reissner--Nordstr\"om solution. This implies that we should have:
\begin{equation}
\gamma_{r_a r_a} \rightarrow \left(1 - \dfrac{2M_{RN}}{r_a} + \frac{Q^2}{r_a^2} \right)^{-1} \; ,
\end{equation}
with $\gamma_{r_a r_a}$ the radial metric component in terms of the areal radius $r_a$, and 
where $M_{RN}$ is the total mass of the star and $Q$ its total charge. Solving for $M_{RN}$ we find that:
\begin{equation} \label{masa-rn}
M_{RN} = \lim_{r_a \rightarrow \infty} \left[ \frac{r_a}{2} \left(1 + \frac{Q^2}{r_a^2}
- \frac{1}{\gamma_{r_a r_a}} \right) \right] \; .
\end{equation}
As before, we need to transform this expression in order to be able to use it in terms of the coordinate radius $r$ instead of $r_a$.  Notice that for a general change of radial coordinate we will have:
\begin{equation}
\gamma_{rr} dr^2 = \gamma_{r_a r_a} d r_a^2 \quad \implies \quad
\frac{1}{\gamma_{r_a r_a}} = \frac{1}{\gamma_{rr}} \left( \frac{dr_a}{dr} \right)^2 \; .
\end{equation}
In our case we have $\gamma_{rr} = \mathrm{a} \Psi^4$ and $r_a =  r \Psi^2 \mathper{b}^{1/2}$, so that:
\begin{equation}
\frac{1}{\gamma_{r_a r_a}} = \frac{\mathper{b}}{\mathper{a}} \left[ 1
+ r \left( \dfrac{\partial_r \mathper{b}}{2\mathper{b}} + 2 \: \dfrac{\partial_r \Psi}{\Psi} \right) \right]^2 \; .
\end{equation}
We finally find:
\begin{equation}
M_{RN} = \lim_{r \rightarrow \infty} \left\{ \frac{\left( r \Psi^2 \mathper{b}^{1/2} \right)}{2}
\left[ 1 + \dfrac{Q^2}{r^2 \Psi^4 \mathper{b}} - \frac{\mathper{b}}{\mathper{a}} \left[ 1
+ r \left( \dfrac{\partial_r \mathper{b}}{2\mathper{b}} + 2 \: \dfrac{\partial_r \Psi}{\Psi} \right) \right]^2  \right] \right\} \; .
\label{eq:MassRN}
\end{equation}
We call this the ``Reissner--Nordstr\"om mass'' in order to distinguish it from the ``integrated mass'' $M_{int}$ defined above. Both our definitions for the total mass should coincide asymptotically. However, in practice the expression for $M_{RN}$ is far more convenient since it gives us the correct mass in the electro-vacuum region, and since the Proca field decays exponentially with $r$ this implies that $M_{RN}$ will also converge exponentially.  In contrast, the integrated mass $M_{int}$ converges quite slowly since the electric field also contributes to the energy density and it decays only as $1/r$.

\vspace{5mm}

%%%%%%%%%%%%%%%%%%%%%%%%%%%%%%%%%%

Finally, we introduce the total binding energy of the star $E_B$.  This is defined as the difference between the total mass--energy $M$ and the rest mass--energy $mN$, that is:
\begin{equation} \label{eq-Eb}
E_B:= M - mN = M - \left( m/q \right) Q \; .
\end{equation}
If the binding energy is negative the star is gravitationally bound, otherwise it is unbound. In the latter case, we expect the configuration to be always unstable against perturbations.

%%%%%%%%%%%%%%%%%%%%
%%%   HORIZONS   %%%
%%%%%%%%%%%%%%%%%%%%

\section{Apparent horizon and horizon mass}

As we will see below, unstable charged Proca stars can collapse to a black hole. In this case we must observe the formation of an apparent horizon, that is, a two-dimensional surface on a spatial slice where the expansion of outgoing null geodesics vanishes. When cosmic censorship and the strong energy condition are verified, the existence of an apparent horizon implies the existence of an event horizon exterior to it. In spherical symmetry, the location of an apparent horizon $r_H$ satisfies the algebraic equation (see e.g.~\cite{Alcubierre08a}):
\begin{equation}
\left[ \frac{1}{\Psi^2 \sqrt{\mathper{a}}} \left( \frac{2}{r} + \frac{\partial_r \mathper{b}}{\mathper{b}} + 4 \: \frac{\partial_r \Psi}{\Psi}\right) - 2K^\theta_\theta \; \right]_{r=r_H} = 0 \; ,
\end{equation}
with $K^\theta_\theta$ the mixed angular component of the extrinsic curvature. If the above equation has more than one solution, the location of the apparent horizon will correspond to largest value of $r$ that satisfies it.  Finding an apparent horizon numerically in the case of spherical symmetry is therefore quite trivial, since it only requires one to evaluate the left hand side of the above equation for all values of $r$ and look for places where it changes sign.

When an apparent horizon has been found, we can calculate its area as $A_H = 4 \pi r_a^2 = 4 \pi r_H^2 \Psi^4_H \mathper{b}_H$, where the subindex $H$ indicates that the corresponding quantity should be evaluated at the horizon position. From the area $A_H$ one can then define the so-called irreducible mass as:
\begin{equation}
M_{irr} := \sqrt{\frac{A_H}{16\pi}} \; .
\end{equation}
The name comes from the fact that this is the minimum mass that a black hole can have for a fixed value of its area, and it should always be less than or equal to the total ADM mass of the spacetime. For a charged Reissner--Nordstr\"om black hole the total ADM mass turns out to be given by:
\begin{equation}
M_{ADM} = M_{irr} + \dfrac{Q^2}{4M_{irr}} \; ,
\end{equation}
where $Q$ is the total charge.  Following this, for the case of our dynamical charged Proca stars when an apparent horizon has been found we will define the total horizon mass $M_H$ as:
\begin{equation} \label{mass-horizon}
M_H := M_{irr} +  \dfrac{Q_H^2}{4M_{irr}} \; ,
\end{equation}
where now $Q_H$ will correspond to the charge integral evaluated at the location of the apparent horizon. When a black hole is formed from the gravitational collapse of a perturbed unstable charged Proca star,  we expect this horizon mass to be less than the total initial mass of the spacetime, since some of the initial Proca field may have escaped to infinity.

%%%%%%%%%%%%%%%%%%%%%%%%%%%%%
%%%   NUMERICAL RESULTS   %%%
%%%%%%%%%%%%%%%%%%%%%%%%%%%%%

\section{Numerical Results} 

To perform the simulations we use the OllinSphere code, which is a fully non-linear time evolution code for numerical relativity in spherical symmetry~\cite{Torres_2014,Alcubierre:2010ea,Ruiz:2012jt}, based on the BSSN formulation adapted to spherical symmetry~\cite{Alcubierre:2010is}. The code uses fourth-order finite differences in space, and a method of lines with a fourth-order Runge--Kutta time integrator. Except for some particular cases that we will mention below, we typically use a spatial mesh with width $\Delta r=0.02$, and a Courant parameter $\Delta t/\Delta r = 0.5$ to ensure numerical stability. In order to suppress high-frequency numerical noise we also apply a standard sixth-order Kreiss--Oliger numerical dissipation to our system of evolution equations~\cite{Kreiss73,Alcubierre08a}.

For the simulations we have always taken $m=1$. However, we can rescale our results to any arbitrary value of $m$ using the fact that our system of equations is invariant under the rescaling:
\begin{equation}
    m \rightarrow \lambda m \; , \qquad 
    q \rightarrow \lambda q \; , \qquad
    \omega \rightarrow \lambda \omega \; , \qquad
    r \rightarrow r/\lambda \; , \qquad 
    E \rightarrow \lambda E \; , \qquad
    \mathcal{E} \rightarrow \lambda \mathcal{E} \; ,
\end{equation}
with $\{\varphi, x, \phi, \mathper{a}, \alpha\}$ unchanged.

For the time evolution we solve equations~(\ref{ev-varphi-3+1}-\ref{ev-Em-3+1}), together with those that come from implementing the BSSN formalism for the geometric variables. The gauge conditions considered throughout the evolution are a vanishing shift vector, and the well-known ``1+log'' slicing condition for the lapse evolution~(see e.g.~\cite{Alcubierre08a} and references therein):
\begin{equation}
\partial_t \alpha = - 2 \alpha K \; ,
\end{equation}
with $K$ the trace of the extrinsic curvature as before.

We also monitor the violation of the Hamiltonian and momentum constraints:
\begin{align}
R^{(3)} - K_{ij} K^{ij} + K^2 &= 16 \pi \rho  \; , \\
\nabla_j^{(3)} \left( K^{ij} - \gamma^{ij} K \right)
&= 8 \pi J^i \; . \label{eq:momentum}
\end{align}
as well as the Gauss constraints~\eqref{cons-Ep-3+1} and~\eqref{cons-Em-3+1}, in order to test the numerical accuracy and convergence of our simulations.  For the explicit form of the Hamiltonian and momentum constraints in the BSSN formulation adapted to spherical symmetry see~\cite{Alcubierre:2010is} (in particular equations~(107) and (108) of that reference).

We also need to set boundary conditions. For the spacetime variables we use constraints preserving boundary conditions as described in~\cite{Alcubierre:2014joa}.  On the other hand, for the Proca and Maxwell variables we impose radiative boundary conditions, which assume that far away a given function behaves as an outgoing spherical wave of the form $f(r,t) = h(r-vt)/r$, with $v$ the corresponding wave speed. This implies that:
\begin{equation}
\partial_t f = - v \left( \partial_r f + f/r \right) \; .
\end{equation}
Assuming that the fields propagate at the speed of light far away we will have $v=\alpha/(\mathper{a}^{1/2} \Psi^2)$. We have found that the best well-behaved case is to apply the radiative condition above only to the scalar potentials $(\phi,\varphi)$, and leave the vector potentials $(x,a)$ and electric fields $(\mathcal{E},E)$ to evolve freely all the way to the boundary using one-sided spatial differences. This results is stable evolutions, and the constraints converge well with increased resolution. However, we have found that the boundary conditions do affect the interior evolution in a significant way unless we place the boundaries very far away. Because of this we typically place our numerical boundaries at $r=100$.

%%%%%%%%%%%%%%%%%%%%%%%%%%%%%%
%%%   SUMMARY OF RESULTS   %%%
%%%%%%%%%%%%%%%%%%%%%%%%%%%%%%

\subsection{Summary of results}

\begin{figure}[ht]
    \centering 
    \includegraphics[width=0.8\textwidth]{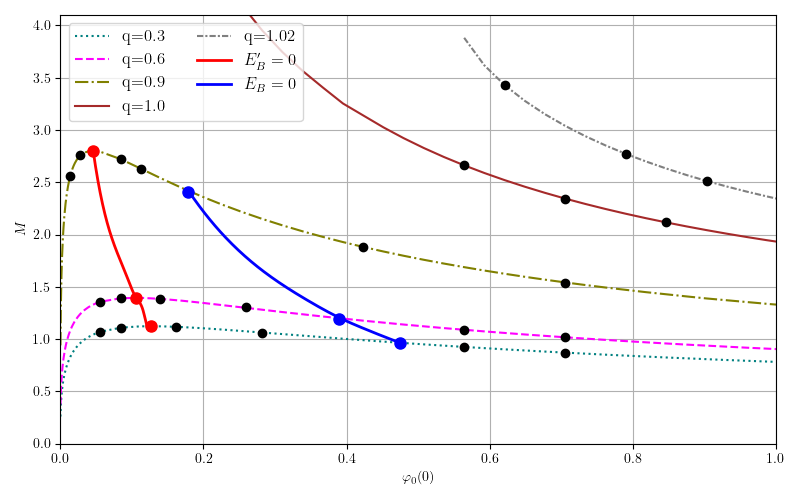} % nombre del archivo
    \caption{Different families of charged Proca star configurations for $q=0.3, 0.6, 0.9, 1.0, 1.02$. The red curve connects the points of maximum mass (minimum binding energy) for each family, while the blue curve connects the points where the binding energy becomes zero. The black dots correspond to the specific configurations considered in Table~\ref{resultado_final} below.}
    \label{masa_phi0_plot}
\end{figure}

The stationary charged Proca star families found in our previous work~\cite{Mio:2025} are shown in Figure~\ref{masa_phi0_plot}, which plots the initial total mass $M$ of the star as a function of the central Proca scalar potential $\varphi_0(0)$ for different values of the charge parameter $q$. Before proceeding, we must clarify that, from now on, for any variable the subscript $0$ denotes its value at the origin $r=0$, while the argument $(0)$ indicates its value at the initial time $t=0$. With this notation in mind, for each charge $q$, there is a whole family of solutions where each value of $\varphi_0(0)$ represents a specific configuration. The profile of the curves is related to the critical value of $q$: for subcritical charges ($q<1$) the mass curve exhibits a rapid initial increase and reaches a well-defined maximum mass whose value depends on $q$.  On the other hand, for supercritical charges ($q>1$) the curve is restricted to a finite interval of values of $\varphi_0(0)$, and is strictly monotonically decreasing throughout.

%El máximo de $q=0.3$ es $0.092$.
%El máximo de $q=0.6$ es $0.076$.
%El máximo de $q=0.9$ es $0.032$.

The red curve on the figure connects the maximum mass of each family, which coincides with the points where the binding energy has a minimum, while the blue curve connects the points where the binding energy is zero ($E_B=0$). These boundaries are important since they divide the parameter space into three regions relevant for subcritical charges. In region I, to the left of the red curve, we find gravitationally bound solutions ($E_B<0$) with have a central potential $\varphi_0(0)$ below the value corresponding to the maximum mass. From similar solutions in the case of uncharged Proca stars, this region is expected to correspond to stable configurations. Region II, located between the red and blue curves, also corresponds to gravitationally bound configurations, but now they have a central potential $\varphi_0(0)$ above the value corresponding to the maximum mass. This region is expected to correspond to unstable but bound configurations, which under perturbations can either collapse to a black hole or possibly migrate to a configuration on the stable region. Finally region III, located to the right of the blue line, corresponds to gravitationally unbound configurations ($E_B>0$) which are expected to always be unstable. In addition, in Figure~\ref{masa_phi0_plot} we also indicate with black dots the specific configurations for which we report the results of our dynamical simulations.  We have performed simulations for many other configurations, but we have chosen these specific ones as examples of the type of behavior that can be found in the different regions of parameter space.

We have evolved both perturbed and unperturbed initial data. In the perturbed case, four different values for the Gaussian perturbation amplitude were considered, $\epsilon = (-0.02,-0.01,+0.01,+0.02)$, with the Gaussian width always taken to be $\sigma = 1$. Notice that in each case we have in fact rescaled the amplitude of the perturbation with the value of the Proca scalar potential at the origin $\varphi_0(0)$, so that a perturbation with e.g. $\epsilon=0.01$ in fact corresponds to a $1\%$ perturbation of the original unperturbed amplitude.

The results of our simulations are summarized in Table~\ref{resultado_final}, where $\epsilon = 0$ corresponds to the unperturbed stationary configurations (which nevertheless in practice are perturbed by numerical truncation error). In the table we report the region of parameter space corresponding to a given configuration, as well as the central values of the Proca potential $\varphi_0(0)$ and $\psi_0(0) \equiv \alpha_0(0) \varphi_0(0)$. For each configuration we also report two frequencies corresponding to the unperturbed solution: the initial eigenvalue $\omega_i$ found by the shooting algorithm by taking $\alpha_0(0)=1$ and $\phi_0(0)=0$, and the final frequency $\omega_f$ found after rescaling the lapse and performing a gauge transformation as detailed in Ref.~\cite{Mio:2025}, which should correspond to the true physical oscillation frequency.  We have chosen to report both values of the frequency in order to make it easier to reproduce our results.

\begin{table}[ht]
    \setlength{\tabcolsep}{7pt} 
    \begin{tabular}{ccccc c ccccc}
        \hline
        & & & & \multicolumn{2}{c}{} & \multicolumn{5}{c}{$\epsilon (\times 10^{-2})$} \\
        \hline
        $q$ & Region & $\varphi_0(0)$ & $\psi_0(0)$ & $\omega_i$ & $\omega_f$ & $\epsilon=-2$ & $\epsilon=-1$ & $\epsilon=0$ & $\epsilon=+1$ & $\epsilon=+2$ \\ 
        \hline
        0.3 & I   & 0.056 & 0.047 & 1.093 & 0.918 & Stable     & Stable     & Stable   & Stable   & Stable \\
        0.3 & I   & 0.085 & 0.066 & 1.131 & 0.900 & Stable     & Stable     & Stable   & Stable   & Stable \\
        0.3 & II  & 0.161 & 0.112 & 1.223 & 0.870 & Migration  & Migration  & Collapse & Collapse & Collapse \\
        0.3 & II  & 0.282 & 0.170 & 1.360 & 0.845 & Migration  & Migration  & Collapse & Collapse & Collapse \\
        0.3 & III & 0.564 & 0.272 & 1.658 & 0.826 & Dispersion & Dispersion & Collapse & Collapse & Collapse \\
        0.3 & III & 0.705 & 0.312 & 1.800 & 0.824 & Dispersion & Dispersion & Collapse & Collapse & Collapse \\
        \hline
        0.6 & I   & 0.056 & 0.045 & 1.087 & 0.930 & Stable     & Stable     & Stable     & Stable   & Stable \\
        0.6 & I   & 0.085 & 0.063 & 1.123 & 0.916 & Stable     & Stable     & Stable     & Stable   & Stable \\
        0.6 & II  & 0.140 & 0.095 & 1.188 & 0.897 & Migration  & Migration  & Migration  & Collapse & Collapse \\
        0.6 & II  & 0.260 & 0.151 & 1.301 & 0.878 & Migration  & Migration  & Migration  & Collapse & Collapse \\
        0.6 & III & 0.564 & 0.254 & 1.641 & 0.858 & Dispersion & Dispersion & Dispersion & Collapse & Collapse \\
        0.6 & III & 0.705 & 0.291 & 1.781 & 0.857 & Dispersion & Dispersion & Collapse   & Collapse & Collapse \\
        \hline
        0.9 & I   & 0.014 & 0.012 & 1.023 & 0.980 & Stable     & Stable     & Stable     & Stable   & Stable \\
        0.9 & I   & 0.028 & 0.021 & 1.040 & 0.972 & Stable     & Stable     & Stable     & Stable   & Stable \\
        0.9 & II  & 0.085 & 0.052 & 1.107 & 0.956 & Migration  & Migration  & Migration  & Collapse & Collapse \\
        0.9 & II  & 0.113 & 0.065 & 1.138 & 0.951 & Migration  & Migration  & Migration  & Collapse & Collapse \\
        0.9 & III & 0.423 & 0.167 & 1.465 & 0.933 & Dispersion & Dispersion & Dispersion & Collapse & Collapse \\
        0.9 & III & 0.705 & 0.234 & 1.742 & 0.931 & Dispersion & Dispersion & Dispersion & Collapse & Collapse \\
        \hline
        1.0 & III & 0.564 & 0.148 & 1.584 & 0.982 & Dispersion & Dispersion & Dispersion & Collapse & Collapse \\
        1.0 & III & 0.705 & 0.177 & 1.720 & 0.980 & Dispersion & Dispersion & Dispersion & Collapse & Collapse \\
        1.0 & III & 0.846 & 0.205 & 1.852 & 0.979 & Dispersion & Dispersion & Dispersion & Collapse & Collapse \\
        \hline
        1.02 & III & 0.621 & 0.127 & 1.633 & 0.999 & Dispersion & Dispersion & Dispersion & Collapse & Collapse \\
        1.02 & III & 0.790 & 0.165 & 1.793 & 0.995 & Dispersion & Dispersion & Dispersion & Collapse & Collapse \\
        1.02 & III & 0.903 & 0.185 & 1.898 & 0.993 & Dispersion & Dispersion & Dispersion & Collapse & Collapse \\
        \hline
    \end{tabular}
    \caption{Results of the simulations for the different configurations indicated by the black dots in Figure~\ref{masa_phi0_plot}.}
    \label{resultado_final}
\end{table}

In general, we find four possible outcomes: (i) Stable: the star just oscillates and remains close to the original stationary state throughout the evolution, which can last for very long times; (ii) Migration: the star changes its configuration and migrates toward a configuration on the stable branch by ejecting some field to infinity; (iii) Collapse: the star collapses to a black hole, and approaches the Reissner--Nordstr\"om solution at late times. (iv) Dispersion: the Proca field disperses to infinity, and the spacetime approaches Minkowski at late times. 

For solutions with subcritical charges $q<1$, we have found that stars in region I are always stable, as expected, while stars in region III are always unstable and may either disperse or collapse to a black hole. Stars in region II are also always unstable, and may either collapse to a black hole or migrate to the stable region. In the case of families with supercritical charges $q \geq 1 $, the end state of the evolutions is invariably either collapse to a black hole or dispersal to infinity. We can also observe that in the case of unstable stars with negative binding energy (region II), there is a tendency towards collapse when the perturbation is positive $\epsilon > 0$, and towards migration to a stable solution when it is negative $\epsilon < 0$. On the other hand, for unstable stars with positive binding energy (region III), a positive perturbation also induces collapse, while a negative perturbation induces dispersion. This is not surprising since a positive perturbation in the scalar potential implies an increase in the initial mass-energy at the center, which causes unstable stars to collapse. Conversely, a negative perturbation implies a decrease in the central mass-energy, which would cause the star to either migrate to a stable less massive star for negative binding energy, or disperse for positive binding energy.

%%%%%%%%%%%%%%%%%%%%%%%%%%%%%%%
%%%   PARTICULAR EXAMPLES   %%%
%%%%%%%%%%%%%%%%%%%%%%%%%%%%%%%

\subsection{Some particular examples} 

\begin{figure}[ht]
\centering
\includegraphics[width=0.8\textwidth]{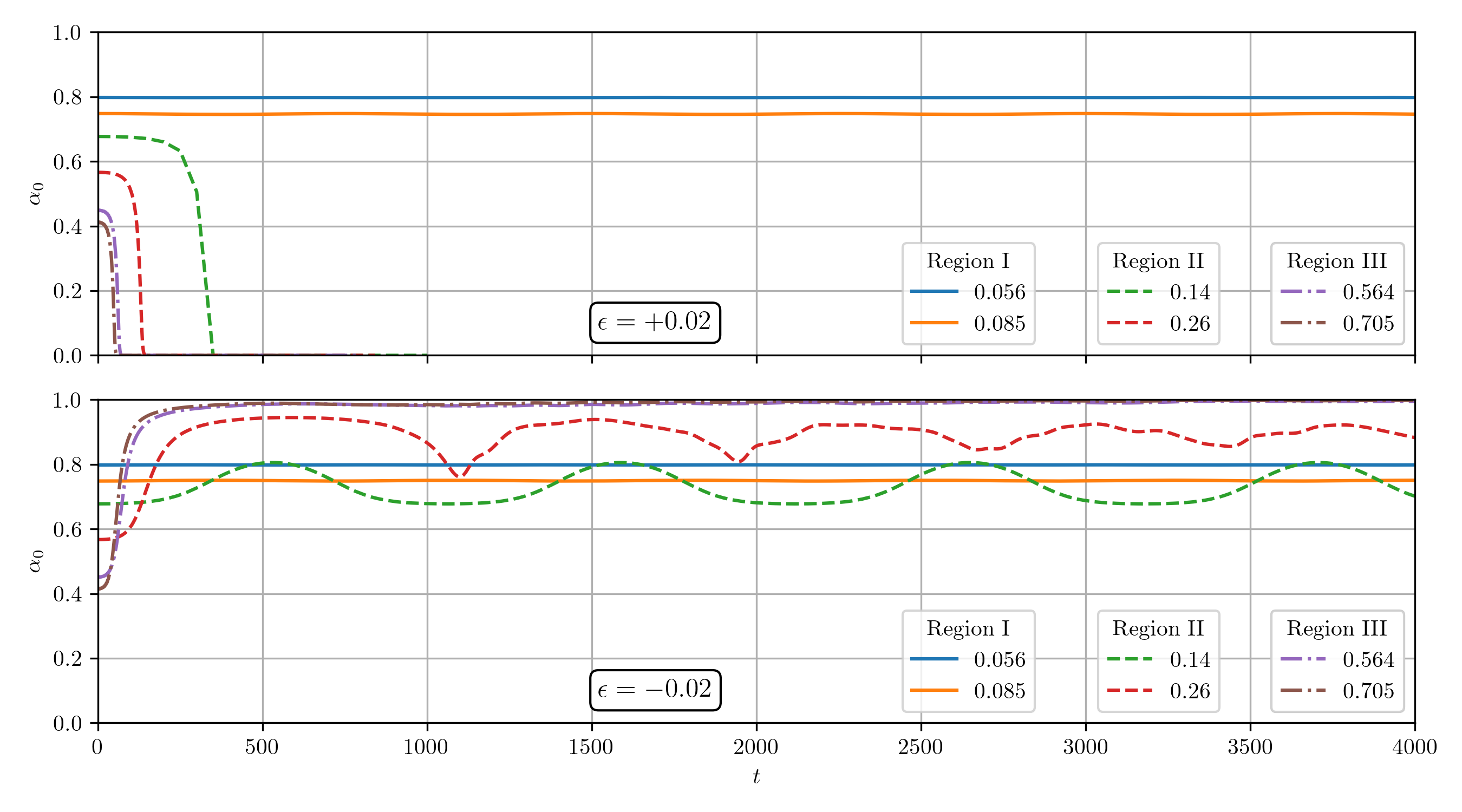} % nombre del archivo
\caption{Time evolution of the central value of the lapse funtcion $\alpha$ for charged Proca star configurations with $q=0.6$, and six different values of the initial central potential $\varphi_0$. The upper panel corresponds to a positive perturbation with amplitude $\epsilon=+0.02$, while the lower panel corresponds to a negative perturbation with $\epsilon=-0.02$.}
\label{alpha_min_q=0.6_def}
\end{figure}

In order to better illustrate the results presented in Table~\ref{resultado_final}, here we present examples of the evolution of  some specific configurations. Figure~\ref{alpha_min_q=0.6_def} shows the evolution of the central value of the lapse function $\alpha$ for the case of charged Proca stars with $q=0.6$ and six different values of the central potential $\varphi_0(0)=(0.056,0.085,0.14,0.26,0.564,0.705)$, which correspond to two configurations in region I, two in region II, and two in region III. First, consider the upper panel of the figure, which shows results for the case when the perturbation has a positive amplitude $\epsilon=+0.02$. We can observe that the configurations in region $I$ remain stable, with the central value of the lapse maintaining essentially constant values (there are in fact very small oscillations around the initial value that are not evident at this scale), while for the configurations in regions II and III the central value of the lapse collapses to zero, indicating the formation of a black hole. The four collapsing configurations do so at different times, with the time of collapse decreasing with larger values of the central potential.

Consider now the lower panel of Figure~(\ref{alpha_min_q=0.6_def}), which shows results for a perturbation with negative amplitude \mbox{$\epsilon=-0.02$}. Again, both configurations in region $I$ are clearly stable with the central value of the lapse remaining constant. The configurations with $\varphi_0(0)=0.140$ and $\varphi_0(0)=0.260$, which correspond to region II, show a more interesting behavior: The central value of the lapse initially grows, and later starts to oscillate around a different larger value. This likely indicates that this configuration is migrating to a more stable configuration in region I. These oscillations, hoewver, last for extremely long times indicating that the migration process is in fact very slow. For the two configurations in region III the central value of the lapse rapidly approaches $\alpha=1$, which corresponds to Minkowski spacetime, indicating that the stars are dispersing.

\begin{figure}[ht]
    \centering
    \includegraphics[width=0.97\textwidth]{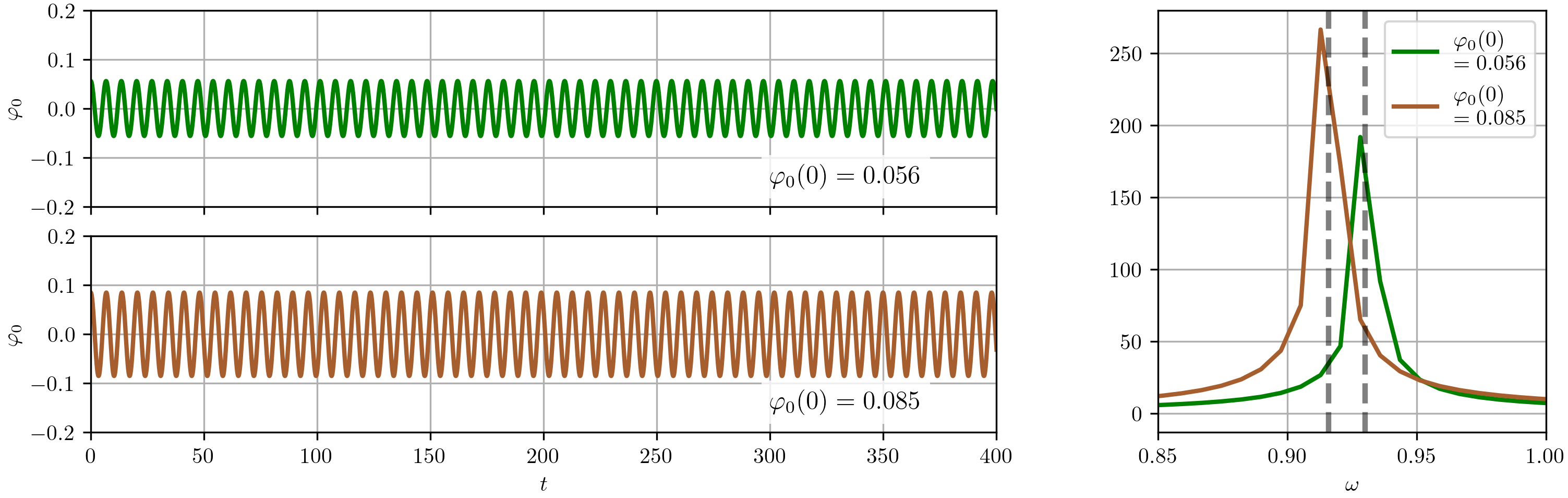} % nombre del archivo
    \caption{Left: Time evolution of the real part of the Proca scalar potential $\varphi_0(t)$ evaluated at the origin, for two unperturbed $(\epsilon=0)$ stable configurations with charge $q=0.6$ and scalar potential $\varphi_0(0)=0.056$ (top panel) and $\varphi_0(0)=0.085$ (bottom panel). Right: Fast Fourier Transform (FFT) of the Proca scalar potential at the origin for the same two configurations.}
    \label{varphi_real_origin_plot_pert=0.00_q=0.6}
\end{figure}

In the left panel of Figure~\ref{varphi_real_origin_plot_pert=0.00_q=0.6} we show the evolution of two unperturbed $(\epsilon=0$) stable configurations with charge $q=0.6$ and central values of the scalar potential $\varphi_0(0)=0.056$ (top panel) and $\varphi_0(0)=0.085$ (bottom panel). The plot shows the time evolution of the real part of the scalar Proca potential $\varphi_0(t)$ evaluated at the origin. The regular oscillations in time reflect the stability of these solutions. In the right panel of the same figure we present the results of a fast Fourier transform (FFT) of $\varphi_0(t)$ for both cases, in order to extract the dominant oscillation frequencies and compare them with the physical frequency $\omega_f$ obtained from the initial data solution given in Table~\ref{masa_phi0_plot}. The peak of the FFT for the case with $\varphi_0(0)=0.056$ is at $\omega=0.928$, which is in close agreement with the expected physical frequency $\omega_f=0.930$, while the peak of the curve for $\varphi_0(0)=0.085$ is at $\omega=0.913$, again in close agreement with the expected frequency $\omega_f=0.916$. This shows that the frequencies $\omega_f$
reported inn the table do indeed corresponf to the physical oscillation frequencies of the stable solutions.

%%%%%%%%%%%%%%%%%%%%%%%%%%%%%%%%%%%%
%%%   MIGRATION AND DISPERSION   %%%
%%%%%%%%%%%%%%%%%%%%%%%%%%%%%%%%%%%%

\subsection{Migration and dispersion}

\begin{figure}[ht]
    \centering
    \includegraphics[width=0.6\textwidth]{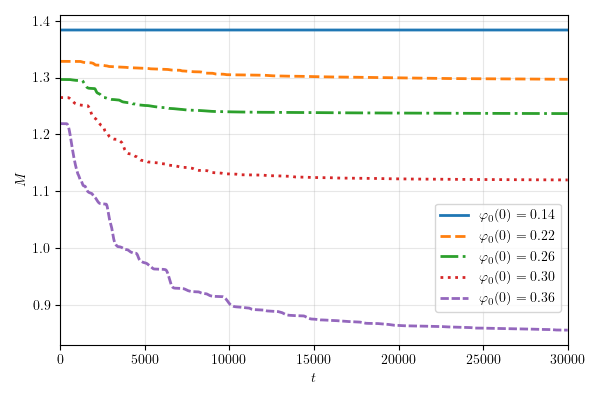}
    \caption{Time evolution of the total Reissner--Nordstr\"om mass evaluated at the grid boundary for five migrating solutions with charge $q=0.6$ and $\varphi_0 = 0.14, 0.22, 0.26, 0.30, 0.36$. These configurations correspond to a negative perturbation with $\epsilon=-0.02$.}
    \label{masa_evol}
\end{figure}

We will now must focus the migrating configurations, specifically those corresponding to the family with charge $q=0.6$ belonging to region II, and subjected to a negative initial perturbation $\epsilon=-0.02$. Here we consider three additional configurations beyond the two already listed in Table~\ref{resultado_final}, corresponding to the initial central potentials \mbox{$\varphi_0=0.14,0.22,0.26,0.30,0.36$}. In Figure~\ref{masa_evol} we show the evolution of the total Reissner--Nordstr\"om mass over time for very long simulations (this time, dr=0.04 was used, ensuring adequate accuracy within a reasonable computational time). In all cases, the mass decreases monotonically and approaches an approximately asymptotic final value (in practice we can not find a true asymptotic value since our numerical methods include dissipative terms). The difference between this asymptotic value and the initial mass corresponds to the energy of the field expelled to infinity during the evolution. It is also evident from the figure that the star expels more mass as the initial central potential $\varphi_0$ increases. Consequently, for the evolution of a low–central-energy star such as the configuration with $\varphi_0(0)=0.140$, the mass variation is imperceptible in the plot.

\begin{figure}[ht]
    \centering
    \includegraphics[width=0.6\textwidth]{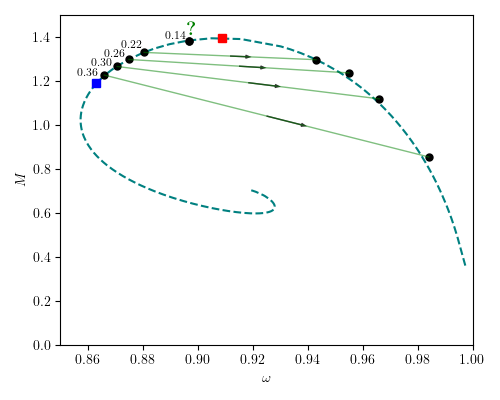}
    \caption{Initial and final states of migrating configuration corresponding to a charge $q=0.6$, perturbation with negative amplitude $\epsilon=-0.02$, and $\varphi_0 = 0.14, 0.22, 0.26, 0.30, 0.36$. The red square indicates the maximum mass configuration, while the black dots represent the initial and final states of our evolving solutions.}
    \label{migrations}
\end{figure}

As just mentioned, the unstable configurations considered here migrate toward stable ones, that is they approach a state that belongs to the parameter space of the stable solutions in region I. In order to determine the final state we measure two parameters that allow us to uniquely specify the corresponding stable solution. Rather than using the pair ($M,\varphi_0(0)$) as in Figure~\ref{masa_phi0_plot}, we instead prefer to consider $M$ and $\omega$, which provides a clearer visualization. After the system has evolved for a sufficiently long time such that the configuration has migrated close to its final state, we determine the total final mass $M$ (although we consider very long simulations with final times $t=30000$, we must keep in mind that the mass is still slowly decreasing). The final frequency $\omega$ is calculated by performing a Fourier transform of the central potential $\varphi_0(t)$ within a suitable late time window, and choosing the dominant frequency. It is important to mention that when doing the Fourier transform one typically finds one dominant peak plus several smaller ones, coming from the fact that the final oscillations are not purely harmonic but are instead modulated.  Presumably, these modulations might disappear after a very long time, but they seem to be extremely long lived.

Our migration results are shown in Figure~\ref{migrations}, which displays the initial and final states for each case. Our procedure to find the final mass and frequency does indeed show that the final configurations correspond to solutions that lie on the stable curve. We must again refer to the particular configuration with $\varphi_0(0)=0.14$, because in this case it was not possible for us to find the final stable configuration. The reason for this is that for this configuration both the mass and frequency do not seem to change significantly during our simulation.  It is likely that much longer numerical evolutions will be required in order to fully determine the final state for this configuration, as the migration proceeds on very long timescales. Nevertheless, the observed trend suggests that this configuration will evolve toward a final state located in the stable branch very close to the maximum of the curve.

\vspace{1mm}

\begin{figure}[ht]
    \centering
    \includegraphics[width=0.97\textwidth]{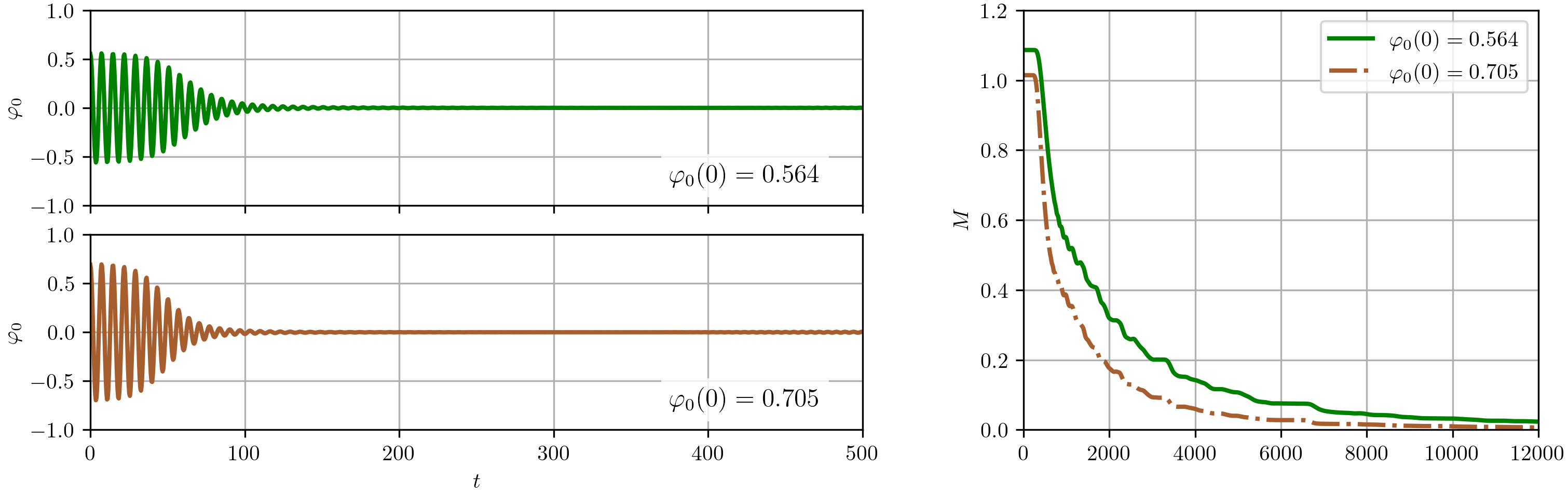} % nombre del archivo
    \caption{Left: Time evolution of the real part of the Proca scalar potential evaluated at the origin $\varphi_0(t)$, for two dispersing configurations with charge $q=0.6$, a negative perturbation with $\epsilon=-0.02$, and initial central values of the scalar potential $\varphi_0(0)=0.564$ (top panel) and $\varphi_0(0)=0.705$ (bottom panel). Right: Time evolution of the Reissner--Nordstr\"om mass for the same two configurations.}
    \label{dispersions}
\end{figure}

Our results also indicate that dispersion processes are restricted to initial configurations located in region III when subjected to negative perturbations. In these cases, the system progressively loses energy until the Proca field disperses completely. This can be seen in Figure~\ref{dispersions}, where we show results from two dispersing configurations with charged parameter $q=0.6$, initial central potentials $\varphi_0(0)= 0.564, 0.705$, and perturbations with $\epsilon=-0.02$. As we can clearly see, the oscillations of the central scalar potential $\varphi_0(t)$ progressively becomes smaller (left panel), while the total mass tends to zero (right panel), indicating that the spacetime approaches Minkowski at late times. In addition, the rate of energy loss increases with the initial central potential of the star.

%%%%%%%%%%%%%%%%%%%%
%%%   COLLAPSE   %%%
%%%%%%%%%%%%%%%%%%%%

\subsection{Collapse}

\begin{figure}[ht]
\centering
\includegraphics[width=0.8\textwidth]{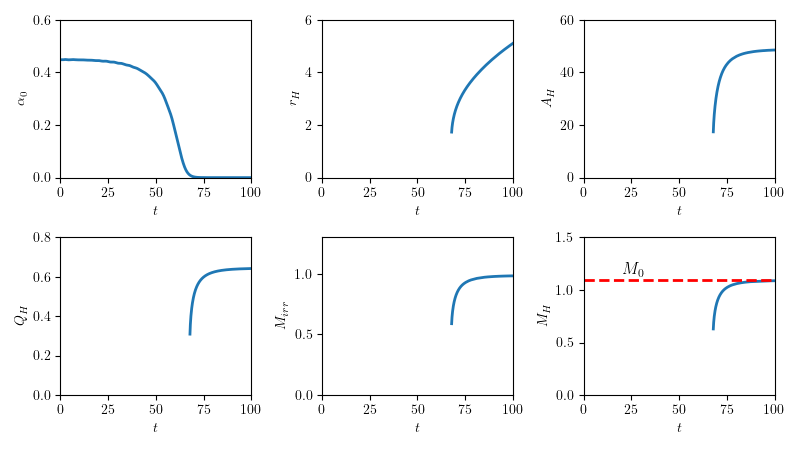} % nombre del archivo
\caption{Evolution of the apparent horizon for a configuration with $q=0.6$, $\varphi_0(0)= 0.564$, and initial positive perturbation $\epsilon=+0.02$. Central value of the lapse $\alpha_0$ (top left), apparent horizon radius $r_H$ (top center), horizon area $A_H$ (top right), charge integral evaluated at the location of the apparent horizon $Q_H$ (bottom left), irreducible mass $M_{irr}$ (bottom center), and total horizon mass $M_H$ (bottom right). In the last panel $M_0$ indicates the initial total mass of star.}
\label{collapse_plot}
\end{figure}

For collapsing solutions, the formation of a black hole is confirmed by monitoring the evolution for the appearance of an apparent horizon. We find that, for all those solutions in which the lapse collapses, an apparent horizon always forms abruptly, confirming black hole formation. As an example, in Figure~\ref{collapse_plot} we show the collapse of the lapse and the appearance of an apparent horizon for a charged Proca star with $q=0.6$, $\varphi_0(0)= 0.564$, and an initial positive perturbation with $\epsilon=+0.02$.
In the figure we show as functions of time: the central value of the lapse $\alpha_0$, the radius of the apparent horizon $r_H$, its area $A_H$, the horizon charge $Q_H$, the horizon irreducible mass $M_{irr}$, and in the final panel a comparison between the final total horizon mass $M_H$ and the initial total mass of the configuration $M_0$ (the initial Reissner--Norstr\"om mass in this case). We can see that for this simulation an apparent horizon suddenly appears at $t \sim 70$, and its final mass is slightly lower that the initial mass. Notice also how the horizon radius $r_H$ grows with time, but its area and mass rapidly approach a constant asymptotic value, showing that the growth of the horizon radius is a purely coordinate effect.

From the uniqueness theorems in the case of spherical symmetry, we know that the final black hole must be of the Reisner--Nordstr\"om class. One way to confirm this is to compare the final asymptotic mass $M_F$ measured far away with the horizon mass $M_H$ inferred from the properties of the black hole horizon given by equation~\eqref{mass-horizon}. For this purpose, we will now consider all those collapsing solutions corresponding to a perturbation with $\epsilon=+0.02$ reported in Table~\ref{resultado_final}. We show the final properties of the apparent horizons for each case in Table~\ref{table-BH-collapse}. The last two columns of the table show the final horizon mass $M_H$ and the final asymptotic Reinsner--Nordstr\"om mass $M_F$.  Notice that these two quantities are calculated in different ways: one from the local horizon properties and one from the asymptotic properties of the spacetime. The final asymptotic mass $M_F$ is taken to be the value corresponding to the final approximately stationary state reached after the collapse, and just before the numerical evolution becomes unreliable due to the growth of numerical errors (however, we must stress the fact that this mass continues to decrease slightly beyond this point).  For all subcritical cases with $q \leq 1$ we can see that we have a very good agreement between both mass calculations. In the supercritical cases with $q=1.02$ the code in fact fails before the masses reach their asymptotic stationary values (remember we have chosen a zero shift vector); this likely explains why $M_F$ does not agree as well with $M_H$ for this cases. For the remaining cases, since $M_H$ coincides with $M_F$ within the numerical uncertainties, we conclude that the black hole formed is indeed of the Reisnner--Nordstr\"om type, as expected.

\begin{table}[ht]
    \centering
    \setlength{\tabcolsep}{8pt} 
    \begin{tabular}{ccccccc}
        \hline
        $q$ & $\varphi_0(0)$ & $A_H$ & $Q_H$ & $M_{\mathrm{irr}}$ & $M_H$ & $M_F$ \\
        \hline
        $0.3$  & $0.161$ & $59.885$  & $0.344$ & $1.092$ & $1.118 \pm 0.006$ & $1.117 \pm 0.003$ \\
        $0.3$  & $0.282$ & $54.227$  & $0.325$ & $1.039$ & $1.064 \pm 0.004$ & $1.064 \pm 0.003$ \\
        $0.3$  & $0.564$ & $41.257$  & $0.276$ & $0.906$ & $0.927 \pm 0.017$ & $0.926 \pm 0.002$ \\
        $0.3$  & $0.705$ & $36.584$  & $0.256$ & $0.853$ & $0.872 \pm 0.014$ & $0.872 \pm 0.002$ \\
        \hline
        $0.6$  & $0.140$ & $77.432$  & $0.847$ & $1.241$ & $1.386 \pm 0.004$ & $1.385 \pm 0.001$ \\
        $0.6$  & $0.260$ & $67.530$  & $0.783$ & $1.159$ & $1.291 \pm 0.008$ & $1.291 \pm 0.002$ \\
        $0.6$  & $0.564$ & $48.840$  & $0.644$ & $0.986$ & $1.091 \pm 0.018$ & $1.090 \pm 0.002$ \\
        $0.6$  & $0.705$ & $42.892$  & $0.594$ & $0.924$ & $1.019 \pm 0.030$ & $1.019 \pm 0.002$ \\
        \hline
        $0.9$  & $0.085$ & $188.697$ & $2.462$ & $1.938$ & $2.720 \pm 0.001$ & $2.719 \pm 0.002$ \\
        $0.9$  & $0.113$ & $177.090$ & $2.377$ & $1.877$ & $2.630 \pm 0.003$ & $2.629 \pm 0.002$ \\
        $0.9$  & $0.423$ & $96.279$  & $1.663$ & $1.384$ & $1.883 \pm 0.009$ & $1.883 \pm 0.003$ \\
        $0.9$  & $0.705$ & $67.497$  & $1.333$ & $1.159$ & $1.542 \pm 0.015$ & $1.542 \pm 0.002$ \\
        \hline
        $1.0$ & $0.564$ & $138.953$ & $2.558$ & $1.663$ & $2.646 \pm 0.011$ & $2.648 \pm 0.012$ \\
        $1.0$ & $0.705$ & $111.600$ & $2.228$ & $1.490$ & $2.323 \pm 0.016$ & $2.334 \pm 0.010$ \\
        $1.0$ & $0.846$ & $93.307$  & $1.988$ & $1.362$ & $2.088 \pm 0.019$ & $2.108 \pm 0.008$ \\
        \hline
        $1.02$ & $0.621$ & $179.303$ & $3.183$ & $1.889$ & $3.230 \pm 0.022$ & $3.341 \pm 0.011$ \\
        $1.02$ & $0.790$ & $130.445$ & $2.596$ & $1.611$ & $2.657 \pm 0.026$ & $2.727 \pm 0.009$ \\
        $1.02$ & $0.903$ & $110.639$ & $2.335$ & $1.484$ & $2.402 \pm 0.026$ & $2.480 \pm 0.008$ \\
        \hline
    \end{tabular}
\caption{Properties of the black holes resulting from the collapsing charged Proca stars of Table~\ref{resultado_final} with a positive perturbation such that $\epsilon=+0.02$.}
\label{table-BH-collapse}
\end{table}

The errors reported in the values of the horizon mass $M_H$ are calculated as follows. From equation~\eqref{mass-horizon} we treat $M_H$ as a function of the irreducible mass and the horizon charge, $M_H = M_H (M_{irr},Q_H)$, and propagate the numerical errors according to:
\begin{equation}
    \delta M_H = \sqrt{\left(\frac{\partial M_H}{\partial M_{irr}} \: \delta M_{irr}\right)^2 + \left(\frac{\partial M_H}{\partial Q_H} \: \delta Q_H\right)^2} 
\end{equation}
where $\delta M_{irr}$ ($\delta Q_H$) is the difference between the irreducible mass (horizon charge) calculated at two different resolutions $\Delta r = 0.02, 0.04$. An analogous procedure was applied to estimate the errors in the final mass $M_F$ using the difference of the values for the two different spatial resolutions.

%%%%%%%%%%%%%%%%%%%%%%%
%%%   CONCLUSIONS   %%%
%%%%%%%%%%%%%%%%%%%%%%%

\section{Conclusions} 

We have performed dynamical evolutions of the charged Proca stars configurations initially found in~\cite{Mio:2025}. By perturbing the initial data in a self-consistent way we have found that three regions can be identified in the parameter space of the solutions, similar to those for the uncharged case $q=0$. Region I, corresponding to configurations with negative binding energy to the left of the maximum mass in Figure~\ref{masa_phi0_plot}, contains all the stable solutions, that is stars that evolve without significant changes over time: The mass and charge remain constant, and the Proca potentials exhibit oscillatory behavior with the same amplitude and a frequency corresponding to that determined by solving the initial eigenvalue problem. In Region II, to the right of the maximum in Figure~\ref{masa_phi0_plot} but still having negative binding energy (to the left of the blue line in  the figure), we find unstable but gravitationally bound. When we evolve these configurations, two different things can happen depending on the specific form of the perturbation: the stars can migrate toward a configuration in the stable branch, or they can collapse into a Reisner--Nordstr\"om black hole. Regions I and II are separated by the curve of solutions with maximum mass (or minimum binding energy). Finally, Region III contains all the unstable and gravitationally unbound solutions with $E_B>0$ (to the right of the blue line in the figure). For such configurations the system evolves toward one of two possible end states: collapse to a black hole or complete dispersion to infinity.

Specifically, the perturbations were implemented by adding or subtracting a small fraction from the central Proca scalar potential of the star. This can be interpreted as adding or removing mass-energy from the star's center. Which of the different outcomes is realized depends on the sign of the perturbation. Thus, a positive initial perturbation for an unstable star implies an increase in the mass–energy, eventually resulting in collapse to a Reissner--Nordstr\"om black hole. In contrast, under a negative perturbation the mass-energy of the star is reduced, either partially (migration) or completely (dispersion). Furthermore, even if we do not add a finite perturbation by hand ($\epsilon=0$), in practice the numerical truncation error induces a perturbation. In this case, however, the final outcome for unstable stars is unpredictable: stars in region II may undergo either migration or collapse, while stars in region III may experience either dispersion or collapse.

Migrations are a very interesting phenomenon. We have been able to verify that a migration is a much slower process than a collapse or a dispersion. A migrating star requires a very long time in order to settle into its final state. However, this timescale decreases as the energy at the center of the star increases. Also, the oscillations of the field in the migrating cases exhibit modulations that seem to remain for extremely long times.  

In~\cite{Mio:2025} we mentioned that, in general, the electric charge plays a dispersing role, which might seem obvious since the electric repulsion opposes the gravitational attraction. In that reference we verified this behavior by observing the radial distribution of matter in the star.  Also, as the charge value increases, more massive stable objects can be found, a fact that is also consistent with the intuitive idea of Coulomb repulsion opposing gravity.

It is important to emphasize here that in this work, both the initial charged Proca star configurations and the implemented perturbations are restricted to spherical symmetry. We believe that the final states of our initial configurations could be different if one considers non-spherical perturbations.  In particular, it is quite possible that the stable configurations presented in this work are not truly stable, and that their dynamic evolution could cause them to migrate to a non-spherical state of lower energy, as has been already shown to be the case for uncharged Proca stars~\cite{Herdeiro:2024a}. We intend to study this possibility in a future work.

%%%%%%%%%%%%%%%%%%%%%%%%%%%
%%%   ACKNOWLEDGMENTS   %%%
%%%%%%%%%%%%%%%%%%%%%%%%%%%

\acknowledgments

The authors wish to thank Carlos Joaquin, Maribel Hernandez and Elisha Candanosa for many useful discussions and comments. This work was partially supported by DGAPA-UNAM project IN100523.  Yahir Mio also acknowledges support from a SECIHTI National Graduate Grant.

%%%%%%%%%%%%%%%%%%%%%%
%%%   REFERENCES   %%%
%%%%%%%%%%%%%%%%%%%%%%

\bibliographystyle{apsrev4-1}
\bibliography{referencias}

%%%%%%%%%%%%%%%%%%%%%%%%%
%%%   END DOCUMENT  %%%%%
%%%%%%%%%%%%%%%%%%%%%%%%%

\end{document}